\documentclass[twocolumn,prl]{revtex4}
\usepackage{graphicx}
\usepackage{amsmath}

\begin{document} 
\title{Relativistic Non-Hermitian Quantum Mechanics}
\author{Katherine Jones-Smith}
\author{Harsh Mathur}
\affiliation{Department of Physics, Case Western Reserve University,10900 Euclid Avenue, Cleveland OH 44106-7079}
\begin{abstract}
We develop relativistic wave equations in the framework of the new non-hermitian ${\cal PT}$ quantum mechanics. 
The familiar Hermitian Dirac equation emerges as an exact result of imposing the Dirac algebra, the criteria of ${\cal PT}$-symmetric quantum mechanics, and relativistic invariance. However, relaxing the constraint that in particular the mass matrix be Hermitian also allows for models that have no counterpart in conventional quantum mechanics. For example it is well-known that a quartet of Weyl spinors coupled by a Hermitian mass matrix reduces to two independent Dirac fermions; here we show that the same quartet of Weyl spinors, when coupled by a non-Hermitian but $\cal{PT}$ symmetric mass matrix, describes a single relativistic particle that can have massless dispersion relation even though the mass matrix is non-zero. 
%
The ${\cal PT}$-generalized Dirac equation is also Lorentz invariant, unitary in time, and CPT respecting, even though as a non-interacting theory it violates ${\cal P}$ and ${\cal T}$ individually. The relativistic wave equations are reformulated as canonical  fermionic field theories to facilitate the study of interactions, and are shown to maintain many of the canonical structures from Hermitian field theory, but with new and interesting new possibilities permitted by the non-hermiticity parameter $m_2$. 
\end{abstract}
\maketitle
In his seminal 1928 paper \cite{dirac}, Dirac proposed that the Hamiltonian for a massive relativistic particle be given by  
\begin{equation}
H_D = - i \boldsymbol{\alpha} \cdot \nabla + \beta, 
\label{eq:diracform}
\end{equation}
where $\boldsymbol{\alpha}$ and $\beta$ are matrices that satisfy the `Dirac algebra'
\begin{equation}
\left\{ \alpha_i, \alpha_j \right \} = 2 \delta_{ij}, 
\hspace{3mm}
\left\{ \alpha_i, \beta \right\} = 0;
 \label{eq:diracalgebra}
\end{equation}
curly brackets denote the anti-commutator. 
To ensure the energy eigenvalues of $H_D$ were real, Dirac assumed $\boldsymbol{\alpha}$ and $\beta$ were Hermitian. If we take $\boldsymbol{\alpha}$ to generate boosts ${\bf K}$ and rotations ${\bf J}$ via  $K_i = i \alpha_i/2$,  $J_i = - i\epsilon_{ijk}\alpha_j \alpha_k/2$
, then by virtue of the Dirac algebra, ${\mathbf J}$ and ${\mathbf K}$ obey the Lorentz algebra, and $H_D$ is Lorentz invariant.  
 Thus Dirac was led to his eponymous equation, which describes relativistic electrons and quarks. Should it turn out to describe neutrinos as well, the Dirac equation would govern all known fermionic matter. 

In this work we consider whether the recently developed formalism of non-Hermitian quantum mechanics can be used to construct $H_D$ 
with matrices that are non-Hermitian but that meet the physically motivated criteria of $\mathcal{PT}$ quantum mechanics, which guarantee real eigenvalues, unitary time evolution, etc \cite{bender,mosta}. 
For fermions time-reversal symmetry is odd, ${\cal T}^2 = -1$ \cite{weinbergqft}, so as a prelude to constructing the non-Hermitian Dirac Hamiltonian we have generalized the formalism of ${\cal PT}$ quantum mechanics to include the case of odd time reversal symmetry\cite{lpu1}.  Fermionic field theories have been considered in the context of ${\cal PT}$ quantum mechanics by Bender {\em et al}~\cite{milton1,milton2,benderjones}; the present manuscript adds to this interesting body of work by explicitly incorporating the $T_{odd}$ character of Dirac fields. 

There have been several important developments in the area of non-Hermitian quantum mechanics in recent years, perhaps most notably in the area of experimental non-linear optics, where an optical analog to the `PT phase transition' has been observed in novel metamaterial structures \cite{expt1,expt2}. 
These and other developments
may advance photonic technology and are of great intrinsic interest \cite{optics1,optics2,optics3, brachisto, mosta, longhi, lots10, graefe, philtransa, jphysa}. But it is worth asking whether the non-Hermitian quantum mechanics remains viable as a fundamental construct, 
as an alternate to the conventional Hermitian framework which underlies equations like the Dirac equation, or whether there is in fact some crucial role played by Hermitian operators in relativistic quantum mechanics. In a similar spirit, \cite{wineland} and \cite{weinberg} have sought to formulate non-linear quantum mechanics and experimentally constrain departures of quantum mechanics from the canonical presumption of linearity. 


Recall the two familiar representations of Eq.(\ref{eq:diracalgebra}) from Hermitian quantum mechanics,  corresponding to Weyl and Dirac fermions. The Weyl representations  $\alpha_i \rightarrow \pm \sigma_i$ are the simplest non-trivial representations of  Eq.(\ref{eq:diracalgebra}) and require the mass matrix $\beta$ be zero,  as no $2 \times 2$ matrix anti-commutes with all three Pauli matrices $\sigma_i$.  The Weyl representations describe a left-handed $(+)$ or right-handed $(-)$ massless fermion; the simultaneous eigenfunctions of $H_D$ and the momentum operator ${\bf p}=-i \nabla$ have the dispersion $E = \pm p$, as is  appropriate for a massless relativistic particle (in units where $c=\hbar=1$). Any other $2 \times 2$ Hermitian representation of Eq.(\ref{eq:diracalgebra}) is unitarily equivalent to one of these. 

The Dirac representation allows for a non-zero mass: choosing the direct sum of a left and right-handed Weyl spinor and the most general form of $\beta$, 
\begin{equation}
\alpha_i \rightarrow \left( 
\begin{array}{cc}
\sigma_i & 0 \\
0 & - \sigma_i
\end{array}
\right), \hspace{5mm}\beta = m \left(
\begin{array}{cc}
0 & 1 \\
1 & 0 
\end{array}
\right)
\label{eq:diracalpha}
\end{equation}
we obtain the celebrated Dirac equation.
The energy and momentum eigenstates have dispersion appropriate for a relativistic particle with mass $m$, $E = \pm \sqrt{ p^2 + m^2 }$, the negative solution having famously led Dirac to propose the existence of antimatter. 

We refer to this choice of $\boldsymbol{\alpha}$ and $\beta$ as the `fundamental' representation of the Dirac algebra Eq.(\ref{eq:diracalgebra}),  because it describes the basic Dirac fermion, {\em i.e.} a pair of  left- and right-handed Weyl spinors coupled by a mass matrix.  Also, the fundamental representation of the massive Dirac fermion is the only representation within the conventional Hermtian theory: any other choice of $\beta$ can be unitarily transformed into this one, and all higher dimensional representations of Eq.(\ref{eq:diracalgebra}) can be decoupled into independent $4\times4$ representations by suitable unitary transformation.

For example, suppose we construct an $8\times8$ representation of the Dirac algebra, a quartet of Weyl spinors:
\begin{equation}
\alpha_i \rightarrow \left[ 
\begin{array}{cccc}
\sigma_i & 0 & 0  & 0 \\
0 & \sigma_i & 0 & 0 \\
0 & 0 & - \sigma_i & 0 \\
0 & 0 & 0 & - \sigma_i
\end{array}
\right].  
\label{eq:quartetalpha}
\end{equation}
In this case the most general choice of mass matrix is
\begin{equation}
\beta = \left( 
\begin{array}{cc}
0 & M \\
M^{\dagger} & 0 
\end{array}
\right) 
\hspace{3mm}
{\rm with} 
\hspace{3mm}
M = \left(
\begin{array}{cc}
m_1 \sigma_0 & m_2 \sigma_0 \\
m_3 \sigma_0 & m_4 \sigma_0 
\end{array}
\right)
\label{eq:quartetbeta}
\end{equation}
where the $m$'s are arbitrary complex numbers and $\sigma_0$ is the $2 \times 2$ identity matrix. But the eigenfunctions of this quartet model have the dispersion $E = \pm \sqrt{ p^2 + \mu_1^2}$ and 
$E = \pm \sqrt{ p^2 + \mu_2^2 }$ where $\mu_1$ and $\mu_2$ are the singular values
of the matrix $M$. So a suitable unitary transformation decouples this $8\times8$ representation into  independent, $4\times4$ fundamental Dirac fermions of mass $\mu_1$ and $\mu_2$ respectively. (See \cite{thesis} for details.) 

We turn now to the variations on the Dirac theory permitted by ${\cal PT}$ quantum mechanics. Our approach is simple: we want to keep the same form of $H_D$ as in Eq. (\ref{eq:diracform}) but relax the constraint that  $\boldsymbol{\alpha}$ and $\beta$ be Hermitian. To ensure real eigenvalues though, we must impose three criteria from ${\cal PT}$ quantum mechanics; these criteria are discussed in detail in \cite{bender, lpu1, thesis} and elsewhere so we just summarize them here. The criteria concern the $\cal{P}$ and $\cal{T}$ operators, which we represent with matrices $S$ and $Z$ respectively, $\mathcal{P} \psi ( r ) = S \psi(-r)$ and $\mathcal{T} \psi(r) = Z \psi^\ast(r)$. In ${\cal PT}$ quantum mechanics the $\cal{P}$ and $\cal{T}$ operators are used to define the inner product; recall there are infinitely many ways to define a valid inner product on a Hilbert space, and matrix operators in that space can be self-adjoint with respect to a given inner product though not necessarily equal to their complex conjugate transpose. 
%
The most natural way to use $\mathcal{P}$ and $\mathcal{T}$ to define the inner product of two wave functions $( \psi, \phi )_{{\cal PT}} = \int d {\mathbf r} [ ({\cal PT} \psi)({\mathbf r}) ]^T Z \phi({\mathbf r})$
can generally leave some states with negative norm so another operator $\cal{C}$ is invoked to flip the sign of those states. The $\cal{CPT}$ inner product replaces the standard Hermitian one; operators are self-adjoint and probability is conserved with respect to this inner product. 

Bender {\em et al} \cite{bender} have found that a Hamiltonian $H$ will have real eigenvalues provided it meets the following criteria:  $[H,\mathcal{PT}]= 0$, $H$ has `unbroken' $\cal{PT}$ symmetry, and $H$ is self-adjoint under the $\cal{PT}$ inner product. So in constructing the ${\cal PT}$-Dirac equation, we assume only that the Dirac algebra and the above criteria are satisfied; $\boldsymbol{\alpha}$ and $\beta$ are not required to be Hermitian and in fact we leave the form of $\boldsymbol{\alpha}$, $\beta$, $S$, and $Z$ undetermined at first, and allow the principles of special relativity and the criteria of ${\cal PT}$ quantum mechanics to determine the exact form these matrices take. It turns out this is enough to ensure the theory is relativistically invariant, and yet departs in marked ways from the Hermitian theory. 
In order for $\cal{P}$ and $\cal{T}$ to be compatible with boosts and rotations we require
$Z \alpha_i^{\ast} = - \alpha_i Z$ and $\{ S, \alpha_i \} = 0$.  Imposing $[\cal{P}, \cal{T}]=$0 requires
$S Z = Z S^{\ast}$; since $\mathcal{T}$ is odd so $Z Z^{\ast} =-1$. For $[H_D, \cal{PT}]=$ 0 we need
$\alpha_i S Z= S Z \alpha_i^{\ast}$ and $\beta S Z = S Z \beta^{\ast}$. Finally, self-adjoincy of $H_D$  
requires
$\alpha_i = Z^T \alpha_i^T Z^{\dagger}$ and $\beta = - Z^T \beta^T Z^{\dagger}$.

Now we are equipped to construct representations of the $\mathcal{PT}$-Dirac equation. 
Naturally we first examine the algebra satisfied by the $\boldsymbol{\alpha}$
matrices since these are no longer explicitly required to be Hermitian. It is easy to show that all
$2 \times 2$ representations of the algebra $\{\alpha_i, \alpha_j\}=0$ are of the left handed type $\alpha_i \rightarrow V \sigma_i V^{-1}$
or the right handed type $\alpha_ i \rightarrow - V \sigma_i V^{-1}$ where $V$ is an invertible matrix; 
note that for Hermitian representations, which form a subset, $V$ must be unitary. This is not a new type of Weyl fermion, however, because it is related to the Hermitian representation by a similarity transformation. 

Next we build the non-Hermitian, $\mathcal{PT}$-analog of the $4\times4$ fundamental representation of Dirac fermion. We call this `Model 4'.  We choose 
$\alpha$ to be a direct sum of  the left- and right-handed representations, 
\begin{equation}
\alpha_i = \left(
\begin{array}{cc}
V \sigma_i V^{-1} & 0 \\
0 & - W \sigma_i W^{-1}
\end{array}
\right)
\label{eq:model4alpha}
\end{equation}
where $V$ and $W$ are invertible matrices.  This is technically a more general  $\alpha$ than in the fundamental Dirac representation.  But after enforcing the criteria enumerated above, we are left with precisely the same form for both $\alpha$ and $\beta$ as in the fundamental Dirac representation, i.e. Eq. (\ref{eq:diracalpha}).  In fact,  Model 4 is equivalent to the fundamental representation in every regard,  right down to the inner product. It is interesting that these cornerstones of Hermitian relativistic quantum mechanics, the Dirac equation and Dirac fermion, do not require a Hermitian Hamiltonian: provided the Dirac algebra is satisfied,  we arrive at exactly the same theory just by imposing constraints from special relativity, parity, and time reversal symmetry. 

But in the $8\times8$ representation we find distinctly new physics. 
For `Model 8' we construct $\alpha_i$ via the direct sum of left-handed and right-handed non-Hermitian $2\times2$ representations:
\begin{equation}
\alpha_i \rightarrow \left[ 
\begin{array}{cccc}
V \sigma_i V^{-1} & 0 & 0  & 0 \\
0 & V \sigma_i V^{-1} & 0 & 0 \\
0 & 0 & -W \sigma_i W^{-1}& 0 \\
0 & 0 & 0 & - W \sigma_i W^{-1}
\end{array}
\right].  
\label{eq:8alpha}
\end{equation}
and we apply the conditions, as in Model 4. This results $\boldsymbol{\alpha}$ identical to that of the  Weyl quartet, Eq.(\ref{eq:quartetalpha}),
but in this case the mass matrix $\beta$ has the more general form
\begin{eqnarray}
\beta & = & \left(
\begin{array}{cc}
0 & M \\
M^{\ast} & 0 
\end{array}
\right)
\hspace{1mm}
{\rm with}
\nonumber \\
M & = & \left[
\begin{array}{cc}
(m_0 + m_3) \sigma_0 & (m_1 - i m_2) \sigma_0 \\
(m_1 + i m_2) \sigma_0 & (m_0 - m_3) \sigma_0 
\end{array}
\right],
\label{eq:model8beta}
\end{eqnarray}
where now the $m's$ are real numbers.   For non-zero $m_2$,  $\beta \neq \beta^\dagger$ and the mass matrix is non-Hermitian. This is a new type of mass matrix and is of course not allowed within the Hermitian theory. 
Even the simplest non-trivial case has interesting features: consider the restricted case where  $m_1=m_3=0$. For the eigenvalues to be real we require $m_0^2 \geq m_2^2$; 
for a given momentum ${\bf p}$ there are four eigenvectors with positive energy
and four with negative energy (just as there are for the Weyl quartet).  The dispersion relation for all eight eigenvectors is 
\begin{equation}
E = \pm \sqrt{ p^2 + m_{{\rm eff}}^2 }
\label{e8}
\end{equation}
where 
\begin{equation} 
m_{{\rm eff}} = \sqrt{ m_0^2 - m_2^2 }
\label{meff}
\end{equation} 
corresponding to a relativistic particle of mass $m_{{\rm eff}}$. 

However this is no ordinary relativistic particle. If  $m_0=m_2$ the restricted Model 8 is the Hamiltonian of an 8-component non-Hermitian, fermion  with two distinct helicity states, and a non-zero mass matrix but a zero effective mass in the dispersion relation. It is worth pointing out that this is an entirely new type of fermion, distinct from the familiar Dirac and Majorana fermions. 

It is interesting to consider these characteristics of Model 8 in the context
of the Standard Model. In the Standard Model we regard quarks and leptons
as massless Weyl fermions that are coupled in a gauge invariant way to
the Higgs field. Effectively this leads to the Weyl fermions being
coupled to one another by a {\em hermitian} mass matrix. With quarks
it is convenient to work with a representation wherein the mass matrices
are diagonal and hence the quarks may be regarded as independent Dirac fermions 
with well defined masses. The price one pays for this is that different
generations of quarks are coupled to the W-boson via the CKM matrix.
With leptons the situation is rather different. It is preferable to
pass to a representation wherein the electron, muon and tau are each
Dirac fermions with a well defined mass, and there is no direct coupling
between different generations of leptons, but the mass matrix of
the three generations of
neutrinos is not diagonal. Within the Standard Model this is
the explanation of the observed phenomenon of flavor oscillation.
It is evident that a mass matrix of the non-hermitian form allowed
by ${\cal PT}$ quantum mechanics might lead to a different phenomenology
of neutrino oscillations and hence the ${\cal PT}$ fermion has potential
relevance to neutrino physics. But we note that it also may have relevance
to quark physics since the CKM matrix is merely a different way of representing
a non-diagonal mass matrix and ${\cal PT}$ quantum mechanics does allow
forms of the mass matrix forbidden by conventional quantum mechanics.
In this connection it is worth noting that 
recent observations from cosmology actually give conflicting values for the sum of the neutrino masses
\cite{pier1,pier2}.
In future work we plan
to explore whether a non-Hermitian mass matrix like
Eq (\ref{eq:model8beta})) could
describe a neutrino which flavor oscillates but propagates masslessly.

Now we would like to dispatch any concern that perhaps Model 8 is merely an elaborate rewriting of a trivial Hermitian model, namely, a pair of $4 \times4$ Dirac Hamiltonians, each of mass $m_{{\rm eff}}$, assembled into an $8\times 8$ block: such a `Dirac pair' model also has a $E = \pm \sqrt{ p^2 + m_{{\rm eff}}^2 }$ with four positive and four negative energy eigenfunctions 
for a given momentum. 
From the eigenfunctions of the Dirac pair and Model 8 
we can simply construct a transformation
to map the Model 8 Hamiltonian to the Dirac pair Hamiltonian, and 
Model 8 wave-functions $\psi_8 ({\mathbf r})$ to Dirac pair wave-functions $\psi_{{\rm Dirac}} ({\mathbf r})$
via the convolution
\begin{equation}
\psi_{{\rm Dirac}} ({\mathbf r}) = \int d {\mathbf r}'  \hspace{1mm} L( {\mathbf r} - {\mathbf r}' ) \psi_8 ({\mathbf r}').
\label{eq:non-local}
\end{equation}
The kernel $L$ has a range set by the non-hermiticity parameter $m_2$.  That the transformation eq (\ref{eq:non-local}) is non-local shows clearly that Model 8 and the Dirac pair model have different physics: if coupled to the same gauge or scalar field
they would yield different outcomes (indeed, it is most relevant to consider $L$ in the context of an interacting theory). However Model 8 also breaks ${\cal P}$ and ${\cal T}$ individually (but respects ${\cal PT}$ by design); see \cite{thesis} for details. 

We can construct Lorentz covariant bilinears to facilitate the study of interactions, although we do not take up any interactions here. 
We write the $8$-component wave-function as a column of four two-component spinors $\xi_1,
\xi_2,
\eta_1,  \textup{and}
\eta_2 $. 
From the form of $\boldsymbol{\alpha}$ for Model 8 we see that the upper two components $\xi_1$ and $\xi_2$ transform
like left handed spinors under boosts and rotations and $\eta_1$ and $\eta_2$ like right-handed.
Furthermore parity exchanges $\xi_1$ with $\eta_1$ and $\xi_2$ with $\eta_2$. 
Thus, just as in Dirac theory, $\xi_i^{\dagger} \eta_j$ and $\eta_i^{\dagger} \xi_j$ are 
all scalars under boosts and rotations. Furthermore the symmetric combination 
$\xi_1^{\dagger} \eta_1 + \eta_1^{\dagger} \xi_1$ is a true scalar being invariant 
under parity whereas the antisymmetric combination $-i (\xi_1^{\dagger} \eta_1 - \eta_1^{\dagger} \xi_1)$
is a pseudo-scalar as it changes sign under parity. Similarly the currents $( \xi_i^{\dagger} \xi_j, \xi_i^{\dagger} 
\boldsymbol{\sigma} \xi_j )$ and $(\eta_i^{\dagger} \eta_j, - \eta_i^{\dagger} \boldsymbol{\sigma} \eta_j)$ are 
four-vectors under boosts and rotations. By making appropriate symmetric and anti-symmetric 
combinations we can construct currents that are true vectors or axial vectors under parity; here again we note the similarity to 
hermitian field theory. Interactions can
now be studied by Yukawa coupling the scalar bilinears to a scalar field or the vector currents to a gauge
field. 

Finally we reformulate Model 8 as a quantum field theory. Define 
the particle and anti-particle creation and annihilation operators as: $c_i({\mathbf p}), c_i^{\star}  ({\mathbf p}),
d_i ({\mathbf p})$ and $d_i^{\star} ({\mathbf p})$ where $i = 1 \ldots 4$. These operators obey the
fermionic anti-commutation relations $
\{ c_i ({\mathbf p}), c_j^{\star} ({\mathbf k}) \} =  \{d_i ({\mathbf p}), d_j^{\star} ({\mathbf k}) \} = \delta ( {\mathbf k} - {\mathbf p} ) \delta_{ij},$
and all other pairs of operators anti-commute. In terms of these creation operators we may write the
Model 8 Hamiltonian as
\begin{equation}
H = \int d {\mathbf p} \sum_{i=1}^4 \sqrt{ p^2 + m_{{\rm eff}}^2 } \left[
c_i^{\star} ( {\mathbf p} ) c_i ( {\mathbf p} ) + d_i^{\star} ( {\mathbf p} ) d_i ( {\mathbf p} ) \right].
\label{eq:fieldhamiltonian}
\end{equation}
Similar expressions can be written for the momentum, parity and other operators.
Next we introduce local field operators 
\begin{eqnarray}
\hat{\psi} ({\mathbf r}) & = & \sum_{i=1}^4 \int d {\mathbf p} \left[ c_i ({\mathbf p}) u_i ({\mathbf p}) e^{i {\mathbf p}
\cdot {\mathbf r}} + d_i^{\star} ({\mathbf p}) v_i ({\mathbf p}) e^{- i {\mathbf p} \cdot {\mathbf r} } \right]
\nonumber \\
\hat{\psi}^{\star} ({\mathbf r} ) & = & 
\sum_{i=1}^4 \int d {\mathbf p} \left[
c_i^{\star} ( {\mathbf p} )  \tilde{u}_i^{\dagger} ({\mathbf p}) e^{- i {\mathbf p} \cdot {\mathbf r} } +
d_i ({\mathbf p}) \tilde{v}_i^{\dagger} ({\mathbf p}) e^{i {\mathbf p} \cdot {\mathbf r} } \right]
\nonumber \\
\label{eq:fieldoperator}
\end{eqnarray}
where we have written the positive energy eigenfunctions of Model 8 as $u_i ({\mathbf p})e^{(i p r)}$ and the negative energy eigenfunctions as $v_i(-{\mathbf p}) e^{i p r}$ where $i = 1, ..., 4$. $\tilde u$ and $\tilde v$ represent the eigenfunctions of $H^{\dagger}_D$.  
The novel twist here is the appearance of $\tilde{u}$ and $\tilde{v}$ in the definition of $\psi^{\star}$. These
operators obey the canonical anti-commutation relation
$\{\psi_a ({\mathbf r}), \psi_b^{\star} ({\mathbf r}') \} = \delta_{ab} \delta( {\mathbf r} - {\mathbf r}' ).$
Thus far we have simply rewritten the non-interacting Model 8 in the language of canonical field theory.
But we may now consistently treat interactions by coupling appropriate bilinears of the field operators
to external scalar and gauge fields. A subtlety that arises in the perturbative analysis of interactions
is that the dynamically determined inner product could also need to be recomputed perturbatively. 
(For earlier attempts in this direction, see for example \cite{shajesh}.)

Non-hermitian quantum mechanics opens up new possibilities for physics beyond the Standard Model. In this Letter we show that ${\cal PT}$ quantum mechanics allows for the existence of fermions of a fundamentally new kind. Among the numerous other directions that deserve further study, we note that the Dirac equation can emerge as the continuum limit of lattice models in solids such as graphene and topological insulators;  whether the non-hermitian Dirac equation might emerge as the low-energy limit in a solid state context is also worth exploration. 

We thank James Bjorken, David Griffiths, and Tom Giblin for helpful comments.

\end{document}